\documentclass[aps,pre,showpacs,twocolumn,amsmath,superscriptaddress]{revtex4-2}
\usepackage{color}
\usepackage{xcolor}
\usepackage[switch*]{lineno}
\usepackage[T1]{fontenc}
\usepackage{graphicx,epsfig,subfigure,dsfont,amssymb,amsmath,amsthm,amsfonts,amsbsy,mathrsfs,amscd}
\usepackage{epstopdf}
\usepackage{array}
\usepackage{booktabs}
\usepackage{amsmath}
\usepackage{bm}
\usepackage{latexsym}
\usepackage{amssymb} %ÓÐ´Ëºê°ü£¬Ôò¿ÉÊ¹ÓÃ\leqslant%
\usepackage{amsthm} %ÕÆ¹ÜÖ¤Ã÷»·¾³£¬ÓÐËüÔò¿ÉÒÔÊ¹ÓÃ\begin{proof}\end{proof}%
\usepackage{times} %ÕÆ¹ÜÎÄ×Ö×ÖÌå£¬ÓÐËüÔòÎªÐÂÂÞÂí×ÖÌå%
\usepackage{graphicx}
\usepackage{tabularx}
\usepackage{enumitem}

\def\1{\mathbf{1}}

\theoremstyle{definition}

\begin{document}
%\linenumbers
%====================================================================================%

\title{Detection of quantum information masking via machine learning}

%====================================================================================%

\author{Sheng-Ao Mao}%毛升奥
\affiliation{School of Science, Hangzhou Dianzi University, Hangzhou
310018, China}
\author{Lin Zhang} %张林
\email{godyalin@163.com}
\affiliation{School of Science, Hangzhou Dianzi University, Hangzhou
310018, China}
\author{Bo Li}%李波
\email{libobeijing2008@163.com} \affiliation{School of Computer and
Computing Science, Hangzhou City University, Hangzhou 310015, China}

\date{\today}

\begin{abstract}
Recently, machine learning has been widely applied in the field of quantum information, notably in tasks such as entanglement detection, steering characterization, and nonlocality verification. However, few studies have focused on utilizing machine learning to detect quantum information masking. In this work, we investigate supervised machine learning for detecting quantum information masking in both pure and mixed qubit states. For pure qubit states, we randomly generate the corresponding density matrices and train an XGBoost model to detect quantum information masking. For mixed qubit states, we improve the XGBoost method by optimizing the selection of training samples. The experimental results demonstrate that our approach achieves higher classification accuracy. Furthermore, we analyze the area under the curve (AUC) of the receiver operating characteristic curve for this method, which further confirms its classification performance.
\end{abstract}

\maketitle


%===================================================%
\section{Introduction}
%===================================================%
Quantum information masking (QIM) is a concept in quantum information processing \cite{PhysRevLett.120.230501,PhysRevA.100.030304,PhysRevLett.126.170505,liu2025experimental}. It refers to the process of using unitary operations to transfer information from a single quantum system into the correlations of a bipartite system, such that the final reduced states of any subsystem are identical. In other words, the subsystems no longer contain any information from the original quantum system. In the seminal work \cite{PhysRevLett.120.230501}, unconditioned masking of all quantum states is deemed impossible. Similar to a series of ``no-go'' theorems in quantum mechanics, such as the no-cloning \cite{wootters1982single}, no-broadcasting \cite{PhysRevLett.100.210502}, no-deleting \cite{kumar2000impossibility}, and no-hiding \cite{PhysRevLett.98.080502} theorems, these are all caused by the linearity (unitarity) properties \cite{PhysRev.28.1049} of quantum theory. Despite the obstacle of universal implementation, QIM admits state-dependent \cite{PhysRevA.100.030304} and probabilistic realizations \cite{PhysRevA.99.052343,PhysRevA.102.022418}. QIM is not only a ``no-go'' theorem in quantum theory but is also closely related to various quantum information processing tasks, including quantum bit commitment \cite{Lie2021quantumonetime} and quantum secret sharing \cite{PhysRevA.100.030304,bai2022quantum}. Therefore, QIM certification is an important task in the field of quantum information.

Machine learning, a core branch of artificial intelligence, is an interdisciplinary field that integrates probability theory, statistics, computer science, and other domains. Its fundamental principle is to build algorithmic models that let computers automatically extract patterns from data, optimize performance, and thus simulate human cognitive learning processes. Based on learning paradigms, machine learning can be categorized into supervised learning, unsupervised learning, semi-supervised learning, and reinforcement learning. These methods have been widely applied in the field of quantum information, such as quantum steering \cite{PhysRevA.100.022314,PhysRevA.104.052427,zhang2020machine,PhysRevA.105.032408}, quantum entanglement classification \cite{PhysRevA.110.042412,chen2021detecting,PhysRevApplied.23.044033}, quantum nonlocality \cite{PhysRevLett.123.190401,PhysRevLett.122.200401,krivachy2020neural,PhysRevA.110.062609,PhysRevLett.120.240402}, spin systems \cite{teoh2020machine,PhysRevX.15.011047,PhysRevB.96.245119}, and quantum phase transition \cite{PhysRevB.100.045129,rem2019identifying,PhysRevA.109.052623}.

Supervised learning is a crucial branch of machine learning. Commonly used supervised learning algorithms include support vector machines (SVM) \cite{cortes1995support}, k-nearest neighbors (KNN) \cite{1053964}, XGBoost \cite{chen2016xgboost}, random forests\cite{10.1023/A:1010933404324}, and neural networks \cite{726791}. Among these algorithms, tree-based models such as XGBoost and random forests demonstrate superior performance in handling structured data \cite{10.5555/3600270.3600307}. XGBoost, in particular, is more efficient in processing large-scale datasets and complex models, while also performing well in preventing overfitting and enhancing generalization ability. Consequently, it has been applied in many fields, including financial market trading \cite{nobre2019combining}, image classification \cite{ren2017novel}, and accelerated Monte Carlo event generation \cite{PhysRevD.107.L071901}.

Recently, supervised machine learning methods, such as neural networks and SVM, have been widely applied in quantum entanglement and quantum steering verification. However, to the best of our knowledge, little attention has been paid to machine learning-based QIM detection. Since QIM plays a significant role in protecting quantum information security and machine learning serves as a powerful tool for solving classification problems, detecting QIM through machine learning is both significant and urgent.

In this paper, we employ the machine learning techniques to tackle QIM detection problem by recasting it as a learning task, which offers a new light on this problem. Specifically, we apply the XGBoost to deal with the problem of QIM detection for both pure and mixed qubit states. For pure qubit states, we use the XGBoost method to detect QIM, which can achieve high classification accuracy. For mixed qubit states, based on active learning techniques \cite{settles.tr09,10.3115/1218955.1219030,10.5555/1613715.1613855}, we propose an XGBoost approach incorporating hybrid sampling, which achieves higher classification accuracy compared to the random sampling and the RandomForest. Furthermore, we compared our proposed approach with the random sampling and the RandomForest in terms of the area under the curve (AUC) of the receiver operating characteristic curve \cite{brown2006receiver}.

The structure of this paper is as follows. Section~\ref{pure_qubit_states} applies the XGBoost method to detect quantum information masking in pure qubit states. In Sec.~\ref{mixed_qubit_states}, we propose an active learning-based XGBoost approach to detect quantum information masking in mixed qubit states. Finally, we conclude in Sec.~\ref{conclusion}.
\section{Pure Qubit States}
\label{pure_qubit_states}
\subsection{Methods}
An arbitrary pure qubit state $|p\rangle$ can be written as $|p\rangle = \cos \frac{x}{2} |0\rangle + e^{i y} \sin \frac{x}{2} |1\rangle \equiv |(x, y)\rangle$, where $x \in [0, \pi]$ and $y \in [0, 2\pi)$.

Ref. \cite{PhysRevA.100.030304} proposed that an arbitrary spherical circle passing through the point $(x_0, y_0)$ (associated with the pure qubit state $|(x_0, y_0)\rangle$) on the Bloch sphere can be expressed as:
\begin{equation}
	\mathcal{C}_{\varphi}^{\beta}\left(\left|\left(x_{0}, y_{0}\right)\right\rangle\right)=\left\{|(x, y)\rangle: \hbar_{\varphi}^{\beta}(x, y)=\hbar_{\varphi}^{\beta}\left(x_{0}, y_{0}\right)\right\},
\end{equation}
where $\hbar_{\varphi}^{\beta}(x, y)=\cos \beta \cos x-\sin \beta \sin x \cos (y-\varphi),\beta \in[0, \pi), \varphi \in[0,2 \pi).$ $\mathcal{C}_{\varphi}^{\beta}\left(\left|\left(x_{0}, y_{0}\right)\right\rangle\right)$ corresponds to a spherical circle passing through the point $\left(x_{0}, y_{0}\right)$ on the Bloch sphere. And all the states $\mathcal{C}_{\varphi}^{\beta}\left(\left|\left(x_{0}, y_{0}\right)\right\rangle\right)$ can be masked by masker $\mathcal{S}_{\varphi}^{\beta}$. Here $\mathcal{S}_{\varphi}^{\beta}$ is an isometry operator defined as follows.\\
$\mathcal{S}_{\varphi}^{\beta}|0\rangle|b\rangle=|0\rangle|u_{0}\rangle+|1\rangle|u_{1}\rangle,\quad\mathcal{S}_{\varphi}^{\beta}|1\rangle|b\rangle=|0\rangle|v_{0}\rangle+|1\rangle|v_{1}\rangle$,\\
where
\begin{equation}
	\renewcommand{\arraystretch}{1.5}
	\begin{array}{l}
		\left|u_{0}\right\rangle=\frac{\sqrt{2}}{2}\big[\cos \big(\frac{\beta}{2}\big) e^{(\varphi+\pi / 4) i}|0\rangle+\cos \big(\frac{\beta}{2}\big) e^{(\varphi+\pi / 4) i}|1\rangle\big], \\
		\left|u_{1}\right\rangle=\frac{\sqrt{2}}{2}\big[\sin \big(\frac{\beta}{2}\big) e^{(\varphi-\pi / 4) i}|0\rangle-\sin \big(\frac{\beta}{2}\big) e^{(\varphi-\pi / 4) i}|1\rangle\big], \\
		\left|v_{0}\right\rangle=-\frac{\sqrt{2}}{2}\big[\sin \big(\frac{\beta}{2}\big) e^{\pi i / 4}|0\rangle+\sin \big(\frac{\beta}{2}\big) e^{\pi i / 4}|1\rangle\big], \\
		\left|v_{1}\right\rangle=\frac{\sqrt{2}}{2}\big[\cos \big(\frac{\beta}{2}\big) e^{-\pi i / 4}|0\rangle-\cos \big(\frac{\beta}{2}\big) e^{-\pi i / 4}|1\rangle\big].
	\end{array}
\end{equation}

For any given pure qubit state $|(x, y)\rangle$, we obtain the label through the following steps. If $|(x,y)\rangle\in\mathcal{C}_{\varphi}^{\beta}\left(\left|\left(x_{0}, y_{0}\right)\right\rangle\right)$, we consider that the pure qubit state can be masked with respect to $|p_{0}\rangle$ and $\mathcal{S}_{\varphi}^{\beta}$, and thus label it as \(+1\). Conversely, if $|(x,y)\rangle\notin\mathcal{C}_{\varphi}^{\beta}\left(\left|\left(x_{0}, y_{0}\right)\right\rangle\right)$, we label it as 0 since it cannot be masked in this context. To balance the data, we randomly generate $l/2$ pure qubit states labeled as \(+1\) and $l/2$ pure qubit states labeled as 0 for the training set, along with 2000 pure qubit states labeled as \(+1\) and 2000 pure qubit states labeled as 0 for the test set, to construct a QIM classifier for pure qubit states. 

The built-in algorithm we use for this classifier is the XGBoost algorithm, an efficient machine learning algorithm based on the gradient boosting framework. The XGBoost algorithm is widely applied to classification and regression tasks involving structured data. Its basic principle is to iteratively optimize model predictions by integrating multiple weak learners. Additionally, regularization and optimization techniques are employed to enhance performance. The base learners of the XGBoost algorithm can be either linear classifiers or decision trees. In this paper, we use the tree model to detect QIM.

For a given data set with $n$ examples and $m$ features $D = \{(\mathbf{a}_i, {b}_{i})\} \ (i=1,2, \cdots, n,\ \mathbf{a}_i \in \mathbb{R}^m,\ {b}_{i} \in \mathbb{R})$, a tree ensemble model uses $K$ additive functions to predict the output.
\begin{equation}
	\hat{b}_{i} = \phi(\mathbf{a}_i) = \sum_{k=1}^{K} f_k(\mathbf{a}_i), \quad f_k \in \mathcal{F},
\end{equation}
where $\mathcal{F} = \{f_{k}(\mathbf{a}) = w_{q(\mathbf{a})}\}$ $\left(q: \mathbb{R}^{m} \rightarrow N, w \in \mathbb{R}^{N}\right)$  is the space of regression trees. $q$ represents the structure of each tree that maps an example to the corresponding leaf index. $N$ represents the number of leaves in the tree. Each $f_k$ corresponds to an independent tree structure $q$ along with its leaf weights $w$. And $w_i$ represents the score on $i$-th leaf.

The regularized objective function $Obj$ can be expressed as follows:
\begin{equation}
	\label{eq:objective_function} 
	Obj=\sum_{i} loss\left(\hat{b}_{i}, {b}_{i}\right)+\sum_{k} \Omega\left(f_{k}\right)
\end{equation}
Here $\sum_{i} loss\left(\hat{b}_{i}, {b}_{i}\right)$ is the loss function. $\hat{b}_{i}$ is the predicted value and ${b}_{i}$ is the true value. $\Omega(f_k)$ is the regularization term and can be expressed as follows:
\begin{equation}
	\Omega(f_k)=\gamma N+\frac{1}{2} \lambda\|w\|^{2}
\end{equation}
where $\gamma$ is used to constrain the number of leaf nodes and $\lambda$ is the L2 regularization parameter.

The tree ensemble model in Eq.\eqref{eq:objective_function} includes functions as
parameters and is trained in an additive manner. By adding $f_t$ to Eq.\eqref{eq:objective_function}, we can obtain
\begin{equation}
	Obj^{(t)}=\sum_{i=1}^{n} loss\left({b}_{i}, \hat{b}_{i}^{(t-1)}+f_{t}\left(\mathbf{a}_i\right)\right)+\Omega\left(f_{t}\right)
\end{equation}
where $\hat{b}_{i}^{(t)}$ is the predicted value of the $i$-th instance at the $t$-th iteration.

By performing second-order Taylor expansion and removing the constant terms, we obtain the simplified objective function at step $t$.
\begin{equation}
	\begin{aligned}
		\tilde{Obj}^{(t)} &=\sum_{i = 1}^{n} [g_{i} f_{t}(\mathbf{a}_i) + \frac{1}{2} h_{i} f_{t}^{2}(\mathbf{a}_i)] + \Omega(f_{t}) \\
		&=\sum_{j = 1}^{N} [(\sum_{i \in I_{j}} g_{i}) w_{j} + \frac{1}{2} (\sum_{i \in I_{j}} h_{i} + \lambda) w_{j}^{2}] + \gamma N
	\end{aligned}
\end{equation}
where $I_{j}=\left\{i \mid q\left(\mathbf{a}_i\right)=j\right\}$ is the instance set of leaf $j$. $g_{i}=\partial_{\hat{b}^{(t-1)}} loss\left({b}_{i}, \hat{b}^{(t-1)}\right)$ and $h_{i}=\partial_{\hat{b}^{(t-1)}}^{2} loss\left({b}_{i}, \hat{b}^{(t-1)}\right)$ are the first-order and second-order gradient statistics of the loss function, respectively.

For a fixed structure $q(\mathbf{a})$, we can obtain the optimal weight $w_{j}^{*}$ of leaf $j$ and the minimum objective function $\tilde{Obj}_{min}^{(t)}$ by
\begin{equation}
	w_{j}^{*}=-\frac{\sum_{i \in I_{j}} g_{i}}{\sum_{i \in I_{j}} h_{i}+\lambda},
\end{equation}
\begin{equation}
	\label{eq:minimum_objective_function}
	\tilde{Obj}_{min}^{(t)}(q)=-\frac{1}{2} \sum_{j=1}^{N} \frac{\left(\sum_{i \in I_{j}} g_{i}\right)^{2}}{\sum_{i \in I_{j}} h_{i}+\lambda}+\gamma N.
\end{equation}

Eq.\eqref{eq:minimum_objective_function} can be regarded as a scoring function to measure the quality of the tree structure $q$. Due to the inability to enumerate all possible tree structures $q$, we employ a greedy algorithm that starts from a single leaf node and iteratively adds branches to the tree. That is to choose the partition with the smallest objective function value and the highest gain function value. Let $I_{L}$ and $I_{R}$ denote the sets of instances assigned to the left and right child nodes after a split, and $I=I_{L} \cup I_{R}$. Then the corresponding gain function can be expressed as follows:
\begin{equation}
	Gain=\frac{1}{2}\left[\frac{\left(\sum_{i \in I_{L}} g_{i}\right)^{2}}{\sum_{i \in I_{L}} h_{i}+\lambda}\!\!+\!\!\frac{\left(\sum_{i \in I_{R}} g_{i}\right)^{2}}{\sum_{i \in I_{R}} h_{i}+\lambda}\!\!-\!\!\frac{\left(\sum_{i \in I} g_{i}\right)^{2}}{\sum_{i \in I} h_{i}+\lambda}\right]-\gamma
\end{equation}

The XGBoost algorithm includes a large number of hyperparameters, such as ``max\_depth'', ``eta'', ``subsample'', and ``colsample\_bytree'', etc. ``max\_depth'' refers to the depth of a single decision tree (that is, the number of splitting layers of the tree). The larger this value is, the more complex the model will be, and thus it is prone to overfitting. Conversely, the smaller this value is, the simpler the model will be, which may lead to underfitting. ``eta'' is the learning rate, which is used to control the weight of each tree. ``subsample'' and ``colsample\_bytree'' are the sample sampling ratio and the feature sampling ratio respectively, and they are used to control the proportion of randomly selected samples and features when each tree is trained. Similarly, if ``eta'', ``subsample'', or ``colsample\_bytree'' is either too large or too small, the classification accuracy will be reduced. Therefore, in this paper, we employ 5-fold cross-validation and grid search to obtain the optimal values of ``max\_depth'', ``eta'', ``subsample'', and ``colsample\_bytree'' in XGBoost. Additionally, since this paper uses XGBoost for a binary classification task, we set the objective function ``objective'' to ``binary: logistic'' and the evaluation metric ``eval\_metric'' to ``logloss''. Meanwhile, we set the remaining hyperparameters to their default values. All the results presented in this paper were obtained by invoking XGBoost via the scikit-learn API in Python.
\subsection{Numerical results}
We used two evaluation metrics for binary classification in our experiments. The first metric is classification accuracy, and the second metric is the area under the curve (AUC) of the receiver operating characteristic curve. The AUC is obtained by plotting the true positive rate (\(\mathrm{TPR}=\mathrm{TP} /(\mathrm{TP}+\mathrm{FN})\)) against the false positive rate (\(\mathrm{FPR}=\mathrm{FP} /(\mathrm{FP}+\mathrm{TN})\)) at various classification thresholds \cite{brown2006receiver}.

\begin{figure}[t]
	\centering
	\includegraphics[width=1\linewidth]{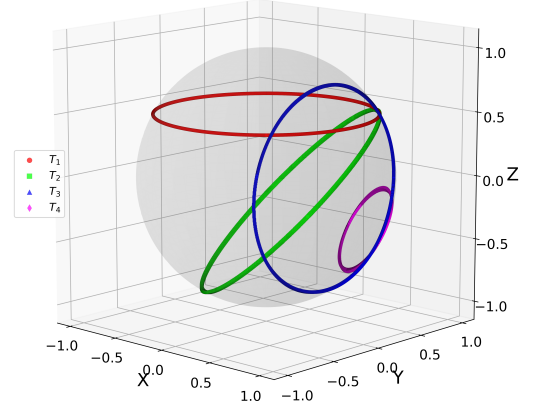}
	\caption{
		Different maskable sets on the Bloch sphere. $T _{1}$, $T _{2}$, $T _{3}$, and $T _{4}$ correspond to the maskable sets $\mathcal{C}_{0}^{0}\left(\left|\left(\frac{\pi }{3} , \frac{\pi }{4} \right)\right\rangle\right)$, $\mathcal{C}_{\frac{\pi }{4}}^{\frac{\pi }{4}}\left(\left|\left(\frac{\pi }{3} , \frac{\pi }{4} \right)\right\rangle\right)$, $\mathcal{C}_{0}^{\frac{\pi }{2}}\left(\left|\left(\frac{\pi }{3} , \frac{\pi }{4} \right)\right\rangle\right)$, and $\mathcal{C}_{\frac{\pi }{3}}^{\frac{\pi }{3}}\left(\left|\left(\frac{2\pi }{3} , \frac{\pi }{5} \right)\right\rangle\right)$ respectively.
	}
	\label{fig:1}
\end{figure}

In the numerical simulation, we randomly select four sets of values for \(x_{0}\), \(y_{0}\), \(\beta\), and \(\varphi\), corresponding to four different maskable sets $\mathcal{C}_{0}^{0}\left(\left|\left(\frac{\pi }{3} , \frac{\pi }{4} \right)\right\rangle\right)$, $\mathcal{C}_{\frac{\pi }{4}}^{\frac{\pi }{4}}\left(\left|\left(\frac{\pi }{3} , \frac{\pi }{4} \right)\right\rangle\right)$, $\mathcal{C}_{0}^{\frac{\pi }{2}}\left(\left|\left(\frac{\pi }{3} , \frac{\pi }{4} \right)\right\rangle\right)$, and $\mathcal{C}_{\frac{\pi }{3}}^{\frac{\pi }{3}}\left(\left|\left(\frac{2\pi }{3} , \frac{\pi }{5} \right)\right\rangle\right)$, which are named $T _{1}$, $T _{2}$, $T _{3}$, and $T _{4}$, respectively. As shown in Fig.~\ref{fig:1}, these maskable sets correspond to distinct spherical circles on the Bloch sphere. For each maskable set $\mathcal{C}_{\varphi}^{\beta}\left(\left|\left(x_{0}, y_{0}\right)\right\rangle\right)$, we generate the dataset using the following procedure:
\begin{itemize}
	\item First, we randomly generate values for \(x\) and \(y\) under the conditions \(x \in [0, \pi]\) and \(y \in [0, 2\pi)\). Since \(|(x, y)\rangle\) represents a pure qubit state, its density matrix is expressed as \(\rho = |(x, y)\rangle \langle (x, y)|\).
	
	\item Since \(\rho\) is a \(2 \times 2\) density matrix, it is enough to use the first element on its diagonal and the real and imaginary parts of the element below the diagonal to construct the feature vector, which is a real vector of three numbers within the interval \((-1, 1)\).
	
	\item For each pure qubit state $|(x,y)\rangle$, if $|(x,y)\rangle\in\mathcal{C}_{\varphi}^{\beta}\left(\left|\left(x_{0}, y_{0}\right)\right\rangle\right)$, we assign it a label \(+1\); otherwise, we assign it a label \(0\).
\end{itemize}

Taking $T _{1}$ as an example, follow the above steps to randomly generate $l/2$ pure qubit states labeled \(+1\) and $l/2$ pure qubit states labeled $0$ as the training set; 2000 pure qubit states labeled as \(+1\) and 2000 pure qubit states labeled as $0$ for the test set. Use the XGBoost algorithm to obtain the prediction accuracy and the area under the curve (AUC). Repeat these steps to obtain six training systems composed of different training sets. Similarly, apply the same procedure for $T _{2}$, $T _{3}$, and $T _{4}$.

\begin{figure}[t]
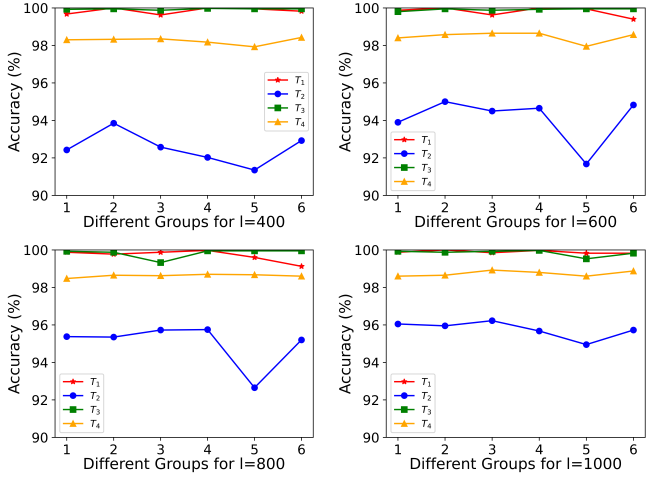

	\centering
	% ???
	\begin{minipage}[t]{0.49\linewidth}
		\centering
		\includegraphics[width=\linewidth]{figure2a}  % ?? 1
	\end{minipage}
	\hfill
	\begin{minipage}[t]{0.49\linewidth}
		\centering
		\includegraphics[width=\linewidth]{figure2b}  % ?? 2
	\end{minipage}
	% ???
	\begin{minipage}[t]{0.49\linewidth}
		\centering
		\includegraphics[width=\linewidth]{figure2c}  % ?? 3
	\end{minipage}
	\hfill
	\begin{minipage}[t]{0.49\linewidth}
		\centering
		\includegraphics[width=\linewidth]{figure2d}  % ?? 4
	\end{minipage}
	\caption{
		The classification accuracy of the XGBoost algorithm across different maskable sets for $l=400$, $600$, $800$, and $1000$. $T _{1}$, $T _{2}$, $T _{3}$, and $T _{4}$ correspond to the maskable sets $\mathcal{C}_{0}^{0}\left(\left|\left(\frac{\pi }{3} , \frac{\pi }{4} \right)\right\rangle\right)$, $\mathcal{C}_{\frac{\pi }{4}}^{\frac{\pi }{4}}\left(\left|\left(\frac{\pi }{3} , \frac{\pi }{4} \right)\right\rangle\right)$, $\mathcal{C}_{0}^{\frac{\pi }{2}}\left(\left|\left(\frac{\pi }{3} , \frac{\pi }{4} \right)\right\rangle\right)$, and $\mathcal{C}_{\frac{\pi }{3}}^{\frac{\pi }{3}}\left(\left|\left(\frac{2\pi }{3} , \frac{\pi }{5} \right)\right\rangle\right)$ respectively. }
	\label{fig:2}
\end{figure}

In Fig.~\ref{fig:2}, we apply the XGBoost algorithm to six different groups of $l$ labeled pure states and obtain the classification accuracy for 4000 unlabeled pure states in the case of $T _{1}$, $T _{2}$, $T _{3}$, and $T _{4}$ (actually these 4000 pure states are labeled, but we treat them as unlabeled data to test the algorithm). When $l = 400$, we observe that for $T_1$, $T_3$, and $T_4$, the classification accuracies all exceed 97\%. For $T_2$, the classification accuracy reaches at least 91.35\%. As the number of training samples $l$ increases, the classification accuracy for $T_1$, $T_3$, and $T_4$ consistently remains above 97\%, while showing significant improvement for $T_2$. Notably, when $l = 1000$, the classification accuracy for $T_2$ reaches at least 94.95\%.

% ?? 4 ??

In Fig.~\ref{fig:3}, we apply the XGBoost algorithm to six different groups of $l$ labeled pure states and obtain the AUC for 4000 unlabeled pure states in the case of $T _{1}$, $T _{2}$, $T _{3}$, and $T _{4}$. As shown in Fig.~\ref{fig:3}, when \( l = 400 \), the AUC exceeds 99\% for \( T_1 \) and \( T_3 \), surpasses 95\% for \( T_2 \), and exceeds 98\% for \( T_4 \). As \( l \) increases, the AUC for \( T_1 \) and \( T_3 \) remains largely unchanged, while the AUC for \( T_2 \) and \( T_4 \) shows slight improvement. Notably, when \( l = 1000 \), the AUC reaches above 98\% and above 99\% for \( T_2 \) and \( T_4 \), respectively. Overall, these results show that the XGBoost algorithm demonstrates strong classification performance in detecting QIM in pure qubit states.

% ?? 4 ??
\begin{figure}[t]
	\centering
	% ???
	\begin{minipage}[t]{0.49\linewidth}
		\centering
		\includegraphics[width=\linewidth]{figure3a}  % ?? 1
	\end{minipage}
	\hfill
	\begin{minipage}[t]{0.49\linewidth}
		\centering
		\includegraphics[width=\linewidth]{figure3b}  % ?? 2
	\end{minipage}
	% ???
	\begin{minipage}[t]{0.49\linewidth}
		\centering
		\includegraphics[width=\linewidth]{figure3c}  % ?? 3
	\end{minipage}
	\hfill
	\begin{minipage}[t]{0.49\linewidth}
		\centering
		\includegraphics[width=\linewidth]{figure3d}  % ?? 4
	\end{minipage}
	\caption{
		The AUC of the XGBoost algorithm across different maskable sets for $l=400$, $600$, $800$, and $1000$. $T _{1}$, $T _{2}$, $T _{3}$, and $T _{4}$ correspond to the maskable sets $\mathcal{C}_{0}^{0}\left(\left|\left(\frac{\pi }{3} , \frac{\pi }{4} \right)\right\rangle\right)$, $\mathcal{C}_{\frac{\pi }{4}}^{\frac{\pi }{4}}\left(\left|\left(\frac{\pi }{3} , \frac{\pi }{4} \right)\right\rangle\right)$, $\mathcal{C}_{0}^{\frac{\pi }{2}}\left(\left|\left(\frac{\pi }{3} , \frac{\pi }{4} \right)\right\rangle\right)$, and $\mathcal{C}_{\frac{\pi }{3}}^{\frac{\pi }{3}}\left(\left|\left(\frac{2\pi }{3} , \frac{\pi }{5} \right)\right\rangle\right)$ respectively. }
	\label{fig:3}
\end{figure}

\section{Mixed Qubit States}
\label{mixed_qubit_states}

\subsection{Methods}
An arbitrary mixed qubit state can be expressed as \(\rho = \left(I_{2} + r_{1} \sigma_{1} + r_{2} \sigma_{2} + r_{3} \sigma_{3}\right)/2 := (r_{1}, r_{2}, r_{3})\), where \(r_{1}^{2} + r_{2}^{2} + r_{3}^{2} \leq 1\) and \(\sigma_{1}\), \(\sigma_{2}\), and \(\sigma_{3}\) are the Pauli matrices. Ref. \cite{PhysRevLett.126.170505} proposed that an arbitrary disk passing through the point $\rho_{0}= (r_{1}^{(0)}, r_{2}^{(0)}, r_{3}^{(0)})$ in the Bloch sphere can be expressed as:
\begin{equation}
\mathcal{D}_{\alpha}^{\theta}\left(\rho_{0}\right)=\{\rho: r_{1} \sin \alpha \cos \theta+r_{2} \sin \alpha \sin \theta+r_{3} \cos \alpha=c\}, 
\end{equation}
where $c=r_{1}^{(0)} \sin \alpha \cos \theta+r_{2}^{(0)} \sin \alpha \sin \theta+r_{3}^{(0)} \cos \alpha, \alpha \in[0, \pi] \text { and } \theta \in[0,2 \pi]$. Furthermore, it has been proven that any qubit disk \(\mathcal{D}_{\alpha}^{\theta}\left(\rho_{0}\right)\) can be masked by the masker \(\mathcal{V}_{\alpha}^{\theta}\). Here \(\mathcal{V}_{\alpha}^{\theta}\) is an isometry defined as follows.
\begin{equation}
	\renewcommand{\arraystretch}{0.8}
	\mathcal{V}_{\alpha}^{\theta}=\left(\begin{array}{cccc}
		\cos (\frac{\alpha}{2}) \!& 0 \!& e^{-i \theta} \sin (\frac{\alpha}{2}) \!& 0 \\
		0 \!& \cos (\frac{\alpha}{2}) \!& 0 \!& e^{-i \theta} \sin (\frac{\alpha}{2}) \\
		0 \!& \sin (\frac{\alpha}{2}) \!& 0 \!& -e^{-i \theta} \cos (\frac{\alpha}{2}) \\
		\sin (\frac{\alpha}{2}) \!& 0 \!& -e^{-i \theta} \cos (\frac{\alpha}{2}) \!& 0
	\end{array}\right) .
\end{equation}

Similar to operations in pure qubit states, for any given mixed qubit state \(\rho\), we obtain the label  through the following steps. If \(\rho \in \mathcal{D}_{\alpha}^{\theta}\left(\rho_{0}\right)\), we consider that the mixed qubit state can be masked with respect to \(\rho_{0}\) and  \(\mathcal{V}_{\alpha}^{\theta}\), and assign it a label of \(+1\). Conversely, if \(\rho \notin \mathcal{D}_{\alpha}^{\theta}\left(\rho_{0}\right)\), we label it as \(0\) since it cannot be masked with respect to \(\rho_{0}\) and \(\mathcal{V}_{\alpha}^{\theta}\). Similarly, we generate the dataset using the following steps.
\begin{itemize}
	\item First, we randomly generate values for \(r_{1}\), \(r_{2}\), and \(r_{3}\) under the condition \(r_{1}^{2} + r_{2}^{2} + r_{3}^{2} \leq 1\). The density matrix \(\rho\) of the mixed qubit state is then obtained using the formula \(\rho = \left(I_{2} + r_{1} \sigma_{1} + r_{2} \sigma_{2} + r_{3} \sigma_{3}\right)/2\).
	\item Similar to the case of pure qubit states, \(\rho\) is a \(2 \times 2\) density matrix. We use the first element on its diagonal and the real and imaginary parts of the element below the diagonal to construct the feature vector, which is a real vector of three numbers within the interval \((-1, 1)\), denoted as \(F\).
	\item For each density matrix \(\rho\), if \(\rho \in \mathcal{D}_{\alpha}^{\theta}\left(\rho_{0}\right)\), it is labeled as \(+1\); if \(\rho \notin \mathcal{D}_{\alpha}^{\theta}\left(\rho_{0}\right)\), it is labeled as \(0\).
\end{itemize}

In real life, unlabeled data is often abundant, while labeled data tends to be scarce. Under such conditions, supervised machine learning often struggles to make accurate predictions on unlabeled data. Therefore, training precise predictive models with limited labeled samples becomes essential. Active learning (AL) techniques provide a solution by selecting the most valuable subset of unlabeled data for labeling, thereby reducing annotation costs \cite{settles.tr09}. This selection of the most valuable data is achieved through query strategy frameworks in active learning. Commonly used query strategies include Uncertainty Sampling, Query-by-Committee, Expected Model Change, and Diversity Sampling, among others.

In this section, we proposed a novel active learning-based XGBoost approach, which we named the AL-XGBoost. We employ a hybrid query strategy that combines uncertainty sampling and diversity sampling to select the most valuable samples for labeling. 

For uncertainty sampling, we use entropy as our uncertainty measure \cite{settles.tr09}:
\begin{equation}
H(\mathbf{x})=-\sum_{A} P\left(A \mid \mathbf{x}\right) \log P\left(A \mid \mathbf{x}\right),
\end{equation}
where \(A\) ranges over all possible labels \(\{0,+1\}\) and \(P\left(A \mid \mathbf{x}\right)\) is the conditional class probability of the class \(A\) for the given unlabeled sample \(\mathbf{x}\).

For diversity sampling, we measure the cosine distance between feature vectors \cite{10.5555/3041838.3041846}:
\begin{equation}
d_{\text{cos}}(F_{i}, F_{j})= 1-\frac{\left\langle F_{i}, F_{j}\right\rangle}{\|F_{i}\|\|F_{j}\|},
\end{equation}
where \(F_{i}\) is the feature vector.

To balance the data, we randomly generate a dataset consisting of 800 mixed states labeled with \(+1\) and 800 mixed states labeled with \(0\) (note that these 1600 mixed states are actually unlabeled, and we need to select the most valuable subset from them for labeling). From this dataset, We randomly select 20 mixed states to label and use them as the initial labeled training set \(\mathcal{L}_0\), while the remaining 1580 mixed states are placed in the initial unlabeled pool \(\mathcal{U}_0\). Additionally, we generate a test set consisting of 2000 mixed states labeled with \(+1\) and 2000 mixed states labeled with \(0\).

\begin{figure}[t]
	\centering
	\includegraphics[width=\linewidth]{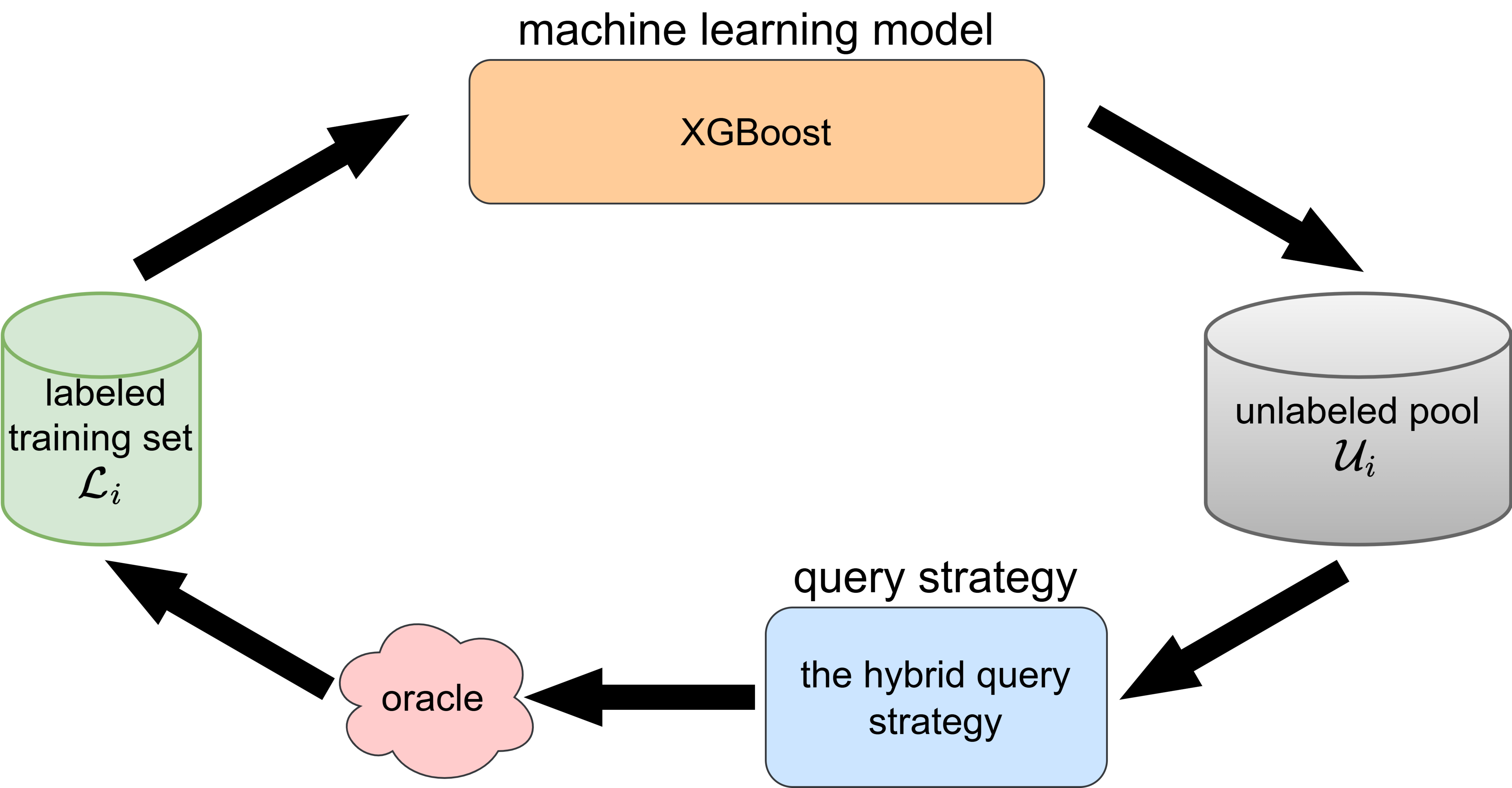}
	\caption{
		An iteration cycle of the AL-XGBoost. Apply the XGBoost algorithm to the labeled training set \(\mathcal{L}_i\) to predict samples in the unlabeled pool \(\mathcal{U}_i\). Then select 5 samples using the hybrid query strategy and present them to the oracle for labeling. Add the queried samples to the labeled training set \(\mathcal{L}_i\) to obtain \(\mathcal{L}_{i+1}\), while simultaneously removing these queried samples from \(\mathcal{U}_i\) to obtain \(\mathcal{U}_{i+1}\). Starting from \(i = 0\), iterate the above steps \( n \) times.
	}
	\label{fig:4}
\end{figure}

Based on the hybrid query strategy, we obtain the training set through the following procedure:
\begin{enumerate}[label=(\arabic*)]
	\item Apply the XGBoost algorithm to the labeled dataset \(\mathcal{L}_i\) (\(i=0,1,2,\ldots\)) to predict the unlabeled mixed states in the pool \(\mathcal{U}_i\) and obtain the class-conditional probability \(P(A \mid \mathbf{x})\) for each unlabeled mixed state belonging to class \(A\).
	
	\item For each unlabeled mixed state, calculate its entropy and select the top 20 unlabeled mixed states with the highest entropy (i.e., the most uncertain). From these 20 unlabeled mixed states, choose the one with the highest entropy and assign it to a set \(\mathit{sp}\), while the remaining 19 are assigned to another set \(\mathit{rp}\).
	
	\item For each point in \(\mathit{rp}\), calculate the sum of its cosine distances to all points in \(\mathit{sp}\). Then, select the point in \(\mathit{rp}\) with the maximum sum of cosine distances, add it to \(\mathit{sp}\), and remove it from \(\mathit{rp}\).

	\item Repeat step (3) until \(\mathit{sp}\) contains five points.

	\item Label the five unlabeled mixed states in \( \mathit{sp} \) (i.e., the most uncertain and diverse unlabeled mixed states from the unlabeled pool \(\mathcal{U}_i\)) and add them to the labeled dataset \(\mathcal{L}_i\) to obtain the updated labeled training set \(\mathcal{L}_{i+1} \). Then, remove these mixed states from the unlabeled pool  \(\mathcal{U}_i\) to obtain \(\mathcal{U}_{i+1} \).
\end{enumerate}
Iterate the above steps \( n \) times to obtain the final labeled dataset \(\mathcal{L}_n\) (composed of \( 5n + 20 \) labeled mixed states) as the training set. By applying the XGBoost algorithm to the test set, we obtain the classification accuracy. The iterative process of the AL-XGBoost is shown in Fig.~\ref{fig:4}. To evaluate the proposed method, we compare it against both the random sampling and the RandomForest. To ensure fair experimental comparisons, we fix all XGBoost hyperparameters as follows:
\begin{itemize}
	\item objective: ``binary: logistic''
	\item eta: 0.09
	\item max\_depth: 5
	\item subsample: 0.9
	\item colsample\_bytree: 1.0
	\item the remaining hyperparameters: default values
\end{itemize}
Meanwhile, the hyperparameters of the RandomForest are fixed as follows:
\begin{itemize}
	\item max\_depth: 5
	\item max\_samples: 0.9
	\item max\_features: 1.0
	\item the remaining hyperparameters: default values
\end{itemize}

\subsection{Numerical results}
\begin{figure}[t]
	\centering
	\includegraphics[width=1\linewidth]{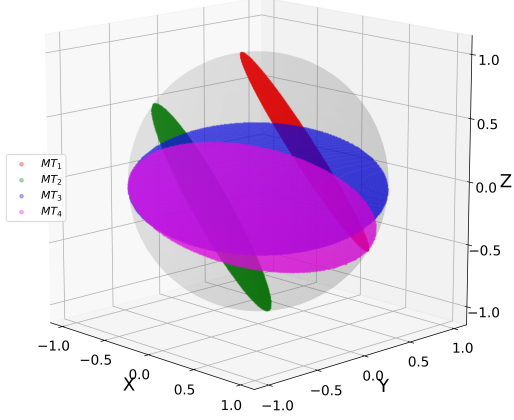}
	\caption{
		Different maskable sets in the Bloch sphere. $MT _{1}$, $MT _{2}$, $MT _{3}$, and $MT _{4}$ correspond to the maskable sets \(\mathcal{D}_{\frac{\pi }{3}}^{\frac{\pi }{3}}\left((\frac{1}{4}, \frac{1}{4}, \frac{1}{4})\right)\), \(\mathcal{D}_{\frac{2\pi }{3}}^{\frac{5\pi }{4}}\left((-\frac{1}{3}, -\frac{1}{2}, \frac{1}{5})\right)\), \(\mathcal{D}_{\frac{\pi }{4}}^{\frac{3\pi }{4}}\left((\frac{1}{3}, \frac{1}{2}, -\frac{1}{4})\right)\), and \(\mathcal{D}_{\frac{3\pi }{4}}^{\frac{5\pi }{3}}\left((-\frac{1}{4}, -\frac{1}{3}, -\frac{1}{5})\right)\) respectively.
	}
	\label{fig:5}
\end{figure}

In the numerical simulation, we randomly select four sets of values for \(r_{1}^{(0)}\), \(r_{2}^{(0)}\), \(r_{3}^{(0)}\), \(\alpha\), and \(\theta\), corresponding to four distinct maskable sets \(\mathcal{D}_{\frac{\pi }{3}}^{\frac{\pi }{3}}\left((\frac{1}{4}, \frac{1}{4}, \frac{1}{4})\right)\), \(\mathcal{D}_{\frac{2\pi }{3}}^{\frac{5\pi }{4}}\left((-\frac{1}{3}, -\frac{1}{2}, \frac{1}{5})\right)\), \(\mathcal{D}_{\frac{\pi }{4}}^{\frac{3\pi }{4}}\left((\frac{1}{3}, \frac{1}{2}, -\frac{1}{4})\right)\), and \(\mathcal{D}_{\frac{3\pi }{4}}^{\frac{5\pi }{3}}\left((-\frac{1}{4}, -\frac{1}{3}, -\frac{1}{5})\right)\), which are named $MT _{1}$, $MT _{2}$, $MT _{3}$, and $MT _{4}$,  respectively. As shown in Fig.~\ref{fig:5}, these maskable sets correspond to different disks in the Bloch sphere.

Taking \( MT_{1} \) as an example, we randomly generated six datasets, each containing 800 mixed states labeled \( +1 \) and 800 mixed states labeled \( 0 \), which are treated as six unlabeled datasets. We also generated a corresponding test set containing 2000 mixed states labeled \( +1 \) and 2000 mixed states labeled \( 0 \). For each unlabeled dataset: 
\begin{itemize} 
	\item For the AL-XGBoost, we first randomly select 20 samples from the dataset for labeling. Next, we apply the hybrid query strategy to select \( l - 20 \) additional samples from the dataset for labeling. Finally, we use all labeled samples as the training set. 
	\item For the random sampling and the RandomForest, we randomly select \( l \) samples from the same dataset, label them, and use the labeled samples as the training set.
\end{itemize} 
Subsequently, the XGBoost algorithm is applied to both the AL-XGBoost and the random sampling to make predictions on the test set, obtaining their respective prediction accuracies and AUC values. For the RandomForest, the random forest algorithm is utilized to make test set predictions and derive corresponding accuracy and AUC metrics. The same procedure is repeated for \( MT_{2} \), \( MT_{3} \), and \( MT_{4} \).

% ?? 4 ??
\begin{figure}[t]
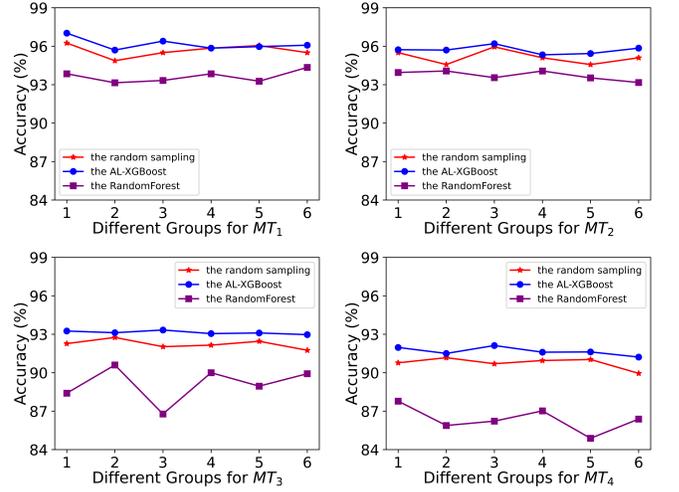

	\centering
	% ???
	\begin{minipage}[t]{0.49\linewidth}
		\centering
		\includegraphics[width=\linewidth]{figure6a}  % ?? 1
	\end{minipage}
	\hfill
	\begin{minipage}[t]{0.49\linewidth}
		\centering
		\includegraphics[width=\linewidth]{figure6b}  % ?? 2
	\end{minipage}
	% ???
	\begin{minipage}[t]{0.49\linewidth}
		\centering
		\includegraphics[width=\linewidth]{figure6c}  % ?? 3
	\end{minipage}
	\hfill
	\begin{minipage}[t]{0.49\linewidth}
		\centering
		\includegraphics[width=\linewidth]{figure6d}  % ?? 4
	\end{minipage}
	\caption{
		The classification accuracy of all three methods across different maskable sets when \(l=1000\). The accuracies by the AL-XGBoost are represented by blue lines with circle, the accuracies by the random sampling are represented by red lines with star, and the accuracies by the RandomForest are represented by purple lines with square, respectively. }
	\label{fig:6}
\end{figure}
In Fig.~\ref{fig:6}, we independently apply the AL-XGBoost, the random sampling, and the RandomForest to select and label 1000 mixed states for each unlabeled dataset, thus obtaining the training sets corresponding to these three methods. We then employ the XGBoost algorithm to both the AL-XGBoost and the random sampling to predict the 4000 mixed states in the test set and obtain the prediction accuracy. For the RandomForest, we employ the random forest algorithm to predict the same 4,000 mixed states and obtain the prediction accuracy. This process is repeated for the maskable sets \(MT_{1}\), \(MT_{2}\), \(MT_{3}\), and \(MT_{4}\). From Fig.~\ref{fig:6}, we observe that for the maskable set \(MT_{1}\), except for group 5 where the AL-XGBoost accuracy is 0.08\% lower than that of the random sampling, the AL-XGBoost achieves higher accuracy, which remains above 95\%. For the maskable sets \(MT_{2}\), \(MT_{3}\), and \(MT_{4}\), the AL-XGBoost consistently achieves higher accuracy, with all exceeding 91\%. Furthermore, across all maskable sets, the accuracy of the AL-XGBoost and the random sampling that employ the XGBoost algorithm is significantly higher than that of the RandomForest which uses the random forest algorithm.

% ?? 4 ??
\begin{figure}[t]
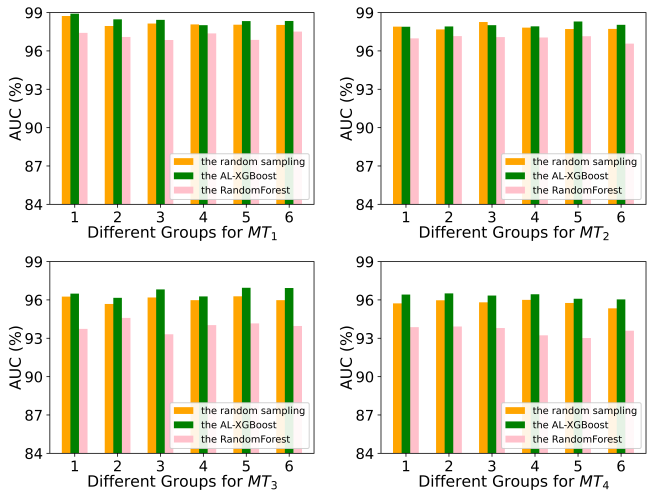

	\centering
	% ???
	\begin{minipage}[t]{0.49\linewidth}
		\centering
		\includegraphics[width=\linewidth]{figure7a}  % ?? 1
	\end{minipage}
	\hfill
	\begin{minipage}[t]{0.49\linewidth}
		\centering
		\includegraphics[width=\linewidth]{figure7b}  % ?? 2
	\end{minipage}
	% ???
	\begin{minipage}[t]{0.49\linewidth}
		\centering
		\includegraphics[width=\linewidth]{figure7c}  % ?? 3
	\end{minipage}
	\hfill
	\begin{minipage}[t]{0.49\linewidth}
		\centering
		\includegraphics[width=\linewidth]{figure7d}  % ?? 4
	\end{minipage}
	\caption{
		The AUC of all three methods across different maskable sets when \(l=1000\). The AUC values by the random sampling are represented by the first column (orange), the AUC values by the AL-XGBoost are represented by the second column (green), and the AUC values by the RandomForest are represented by the third column (pink). }
	\label{fig:7}
\end{figure}
In Fig.~\ref{fig:7}, we independently apply the AL-XGBoost, the random sampling, and the RandomForest to select and label 1000 mixed states for each unlabeled dataset, thus obtaining the training sets corresponding to these three methods. We then employ the XGBoost algorithm to both the AL-XGBoost and the random sampling to predict the 4000 mixed states in the test set and obtain the AUC. For the RandomForest, we employ the random forest algorithm to predict the same 4,000 mixed states and obtain the AUC. This process is repeated for the maskable sets \(MT_{1}\), \(MT_{2}\), \(MT_{3}\), and \(MT_{4}\). As shown in Fig.~\ref{fig:7}, for the maskable set \(MT_{1}\), except for group 4 where the AUC of the AL-XGBoost is 0.06\% lower than that of the random sampling, the AL-XGBoost achieves higher AUC values in all other groups. In the case of \(MT_{2}\), the AUC of the AL-XGBoost is within 0.25\% lower than that of the random sampling in groups 1 and 3, but it demonstrates higher values in the remaining groups. For the maskable sets \(MT_{3}\) and \(MT_{4}\), the AUC of the AL-XGBoost is generally higher than that of the random sampling. Furthermore, across all maskable sets, the AUC of the AL-XGBoost and the random sampling is significantly higher than that of the RandomForest. In summary, both the classification accuracy and AUC of the AL-XGBoost are overall higher than those of the random sampling and the RandomForest. This indicates that the AL-XGBoost can effectively improve the classification performance of information masking in mixed qubit states.

\begin{figure}[t]
	\centering
	\includegraphics[width=\linewidth]{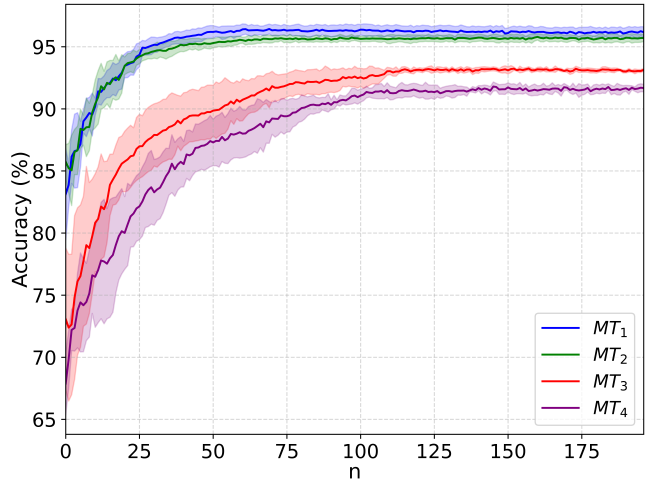}
	\caption{
		When \(l = 1000\), the evolution of average prediction accuracy and its standard deviation for the AL-XGBoost during iterative training across different maskable sets \(MT_{1}\), \(MT_{2}\), \(MT_{3}\), and \(MT_{4}\). The lines and their shaded areas represent the average prediction accuracies and their standard deviations, respectively.
	}
	\label{fig:8}
\end{figure}
From Fig.~\ref{fig:8}, we observe that for all four maskable sets, as the number of iterations \(n\) increases, the average accuracy of the AL-XGBoost initially improves significantly and then tends to stabilize after a certain number of iterations. Additionally, the shaded areas of the AL-XGBoost gradually shrink as \(n\) increases, indicating a reduction in the standard deviations. This shows that the models are well-trained.

\begin{figure*}[t]
	\centering
	\includegraphics[width=\textwidth]{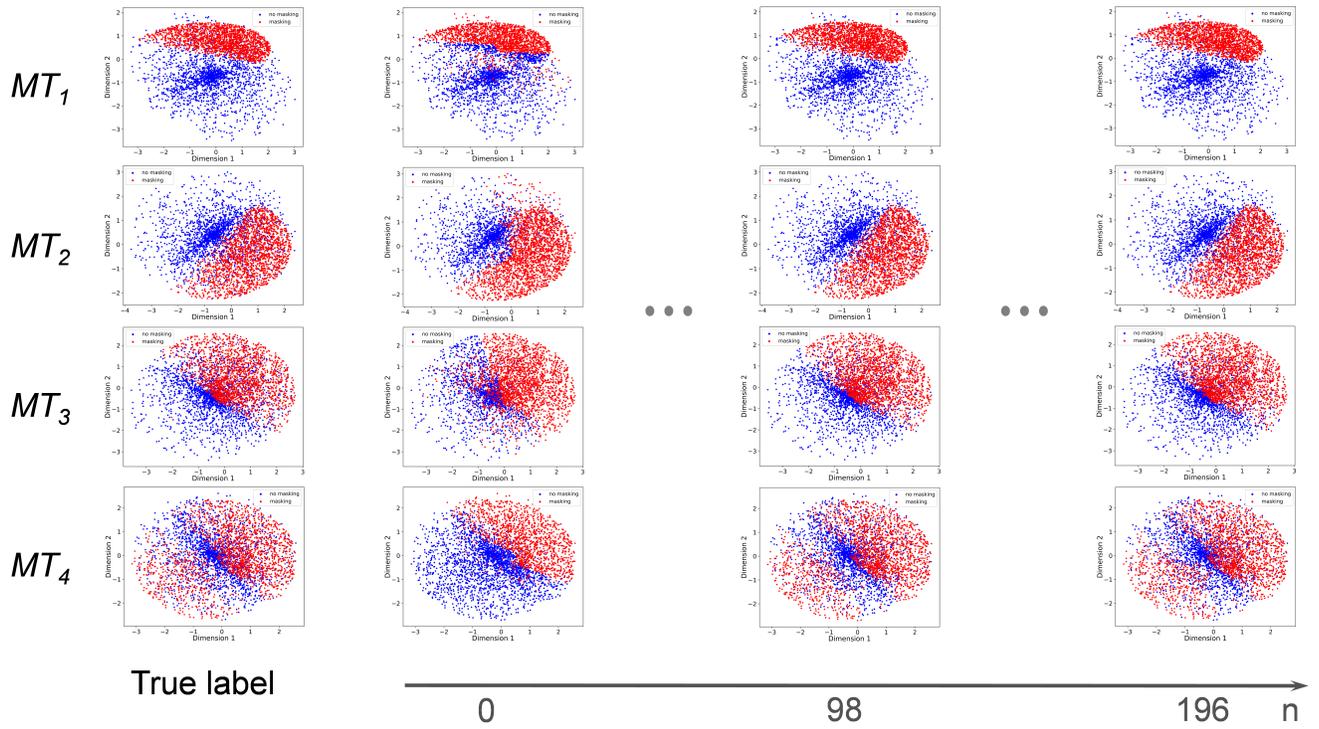}
	\caption{
		True and predicted label distributions of feature vectors in the test set for different maskable sets \(MT_{1}\), \(MT_{2}\), \(MT_{3}\), and \(MT_{4}\) when \(l = 1000\). The column labeled ``True label'' represents the true label distribution of feature vectors in the test set for each maskable set. The remaining three columns represent the predicted label distributions during the iterative training process of the AL-XGBoost.
	}
	\label{fig:9}
\end{figure*}
In Fig.~\ref{fig:9}, we use Principal Component Analysis (PCA) to project the test set’s feature vectors into a two-dimensional space, thereby visualizing the true label distribution and the predicted label distributions of the AL-XGBoost (during its iterative training) for each maskable set. Since the six groups of training systems corresponding to each maskable set have similar training results, we implement the above visualization operation on the first group of training systems for each maskable set. From Fig.~\ref{fig:9}, we observe that for maskable sets \(MT_{1}\), \(MT_{2}\), and \(MT_{3}\), at the initial iteration (\(n = 0\)), the predicted label distributions are broadly similar to the corresponding true label distributions. However, the model tends to misclassify samples near the classification boundaries. By the final iteration, the model can effectively distinguish samples close to the classification boundary. For the maskable set \(MT_{4}\), the predicted label distribution at \(n=0\) differs significantly from the true label distribution. However, as the number of iterations increases, the model's predicted label distribution gradually converges toward the true label distribution.

\section{DISCUSSION AND CONCLUSION}
\label{conclusion}
\begin{figure}[t]
	\centering
	\includegraphics[width=\linewidth]{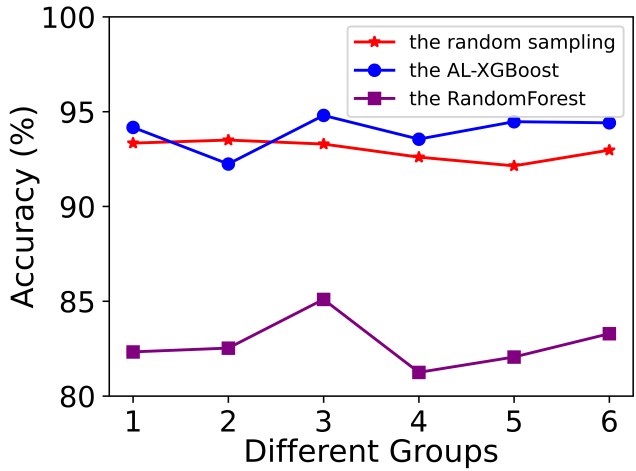}
	\caption{
		When \(l = 2000\), the classification accuracy of all three methods in the four-class experiment. The accuracies by the AL-XGBoost are represented by blue lines with circle, the accuracies by the random sampling are represented by red lines with star, and the accuracies by the RandomForest are represented by purple lines with square, respectively.
	}
	\label{fig:10}
\end{figure}

To evaluate the effectiveness of the proposed method in multi-class classification tasks, we conduct a four-class experiment. The samples in $MT_{1}$, $MT_{2}$, $MT_{3}$, and $MT_{4}$ correspond to classes 0, 1, 2, and 3, respectively. We randomly generate six datasets, each consisting of 1000 samples per class (4000 samples total per dataset). These six datasets serve as our unlabeled datasets. From each unlabeled dataset, we select $l$ samples for labeling to create a training set, thus yielding six distinct training sets for the cases of the AL-XGBoost, the random sampling, and the RandomForest. Concurrently, we generate a corresponding test set containing 2000 samples per class (8000 samples total). We then compare the classification accuracy of the AL-XGBoost, the random sampling, and the RandomForest across these six training setups, as shown in Fig.~\ref{fig:10}. We find that our method can achieve a maximum accuracy of 94.8\%. However, we only consider the four-class scenario. In future work, we consider extending the approach to more complex multi-class cases.

In conclusion, we explored the use of supervised machine learning methods to identify quantum information masking in both pure and mixed qubit states. For pure qubit states, we randomly generated the corresponding density matrices and used the XGBoost to classify whether they were information masking. We found that high classification accuracy can be achieved with few samples on certain maskable sets. Furthermore, for mixed qubit states, we proposed an active learning-based XGBoost approach that selects the most valuable samples as the training set through hybrid sampling. Our numerical simulations demonstrated that this method exhibits stronger classification performance compared to the random sampling and the RandomForest in most cases, thus providing a high-performance solution to QIM detection problem. Although the AL-XGBoost method proposed in this paper improves classification performance, it may require more time than the random sampling since the AL-XGBoost needs to iteratively invoke the XGBoost algorithm multiple times during the selection of training samples. The goal of future research is to detect information masking in more complex states by employing efficient and stable supervised or semi-supervised methods.

\begin{acknowledgments}
This research is supported by Zhejiang Provincial Natural Science Foundation of China under Grant No. LZ23A010005, LZ24A050005; Jiangxi Provincial Natural Science Foundation of China, under Grants No. 20213BCJL22054; and NSFC under Grant No.12175147.
\end{acknowledgments}

%=============================================================================%
\bibliographystyle{iopart-num}
\bibliography{article}

\end{document}